\documentclass[aps,prl,10pt,twocolumn,preprintnumbers,superscriptaddress,showpacs,floatfix,nofootinbib]{revtex4-2}
\usepackage{textcomp,gensymb}
\usepackage{hyperref}
\usepackage{graphicx}
\usepackage{amsfonts,amsmath,amssymb,bm,bbm}
\usepackage{color,soul}
\usepackage[dvipsnames]{xcolor}
\usepackage{slashed}
\usepackage{flushend}
\usepackage{physics}
\usepackage{graphicx}
\usepackage{dcolumn}
\usepackage{float}
\usepackage{comment}
\usepackage{tikz}
\usepackage{subcaption}
\usepackage[normalem]{ulem}
\usepackage[compat=1.1.0]{tikz-feynman}
\usepackage[utf8]{inputenc}

\usepackage{ragged2e}
\usepackage{subcaption}
\usepackage{csquotes}

\hypersetup{
    pdfnewwindow=true,      
    colorlinks=true,       
    linkcolor=blue,          
    citecolor=blue,        
    filecolor=blue,      
    urlcolor=blue        
}
\usepackage{graphicx}

\begin{document}
\preprint{FERMILAB-PUB-26-0020-PPD}
\preprint{MI-HET-876}
\title{Dark-Matter–Induced Nuclear Deexcitation at the Short Baseline Near Detector with Ab Initio Nuclear Theory}
\author{Bhaskar Dutta}
\email{dutta@tamu.edu}
\affiliation{Mitchell Institute for Fundamental Physics and Astronomy, Department of Physics and Astronomy, Texas A\&M University, College Station, Texas 77843, USA}
\author{Debopam Goswami}
\email{debopam22@tamu.edu}
\affiliation{Mitchell Institute for Fundamental Physics and Astronomy, Department of Physics and Astronomy, Texas A\&M University, College Station, Texas 77843, USA}
\author{Baishan Hu}
\email{baishanhu4phys@gmail.com}
\affiliation{Cyclotron Institute and Department of Physics and Astronomy, Texas A$\&$M University, College Station, Texas 77843, USA}
\author{Wei-Chih Huang}
\email{s104021230@tamu.edu}
\affiliation{Mitchell Institute for Fundamental Physics and Astronomy, Department of Physics and Astronomy, Texas A\&M University, College Station, Texas 77843, USA}
\author{Vishvas Pandey}
\email{vpandey@fnal.gov}
\affiliation{Fermi National Accelerator Laboratory, Batavia, Illinois 60510, USA}

\begin{abstract}
We explore the sensitivity of the short-baseline near detector (SBND) experiment to light dark matter using MeV-scale electromagnetic activity. Inelastic scattering of dark matter with argon nuclei can excite nuclear states that subsequently deexcite via the emission of MeV-scale photons, producing localized low-energy ``blip” signatures in a liquid argon time projection chamber. We perform state-of-the-art ab initio nuclear calculations, including all relevant argon excited states with energies up to 18 MeV, to provide reliable predictions for these signals. After accounting for relevant backgrounds, we find that SBND can probe previously unexplored regions of parameter space for light dark matter.
\end{abstract}
\maketitle

{\bf{\emph{Introduction} -}} Proton- and electron-beam–dump experiments provide excellent opportunities to probe light-dark-matter models through both appearance and disappearance channels. Among existing results, ongoing proton-beam–based neutrino experiments, such as COHERENT~\cite{COHERENT:2021pvd} and CCM~\cite{CCM:2021leg}, as well as earlier experiments like LSND~\cite{LSND:2001akn} and MiniBooNE~\cite{MiniBooNE:2017nqe}, have explored regions of sub-GeV dark matter (DM) parameter space by using neutrino detectors to search for DM produced when high-intensity proton beams strike a target. Meanwhile, the electron- and muon-beam–based NA64 e/$\mu$\cite{Andreev:2024lps, Banerjee:2019pds, NA64:2024klw} experiments have probed the parameter space through disappearance channels associated with light-mediator decays.

Recently, it has been shown that light DM can be efficiently probed via DM–nucleus inelastic scattering channels by utilizing deexcitation photons in the context of the ongoing CCM experiment~\cite{Dutta:2023fij, Dutta:2022tav}. Although the inelastic scattering cross section is typically smaller than that of elastic channels, the MeV–range final-state photons benefit from significantly reduced neutrino-induced backgrounds in the relevant unconstrained parameter space. Additionally, the discrete energy spectra of nuclear de-excitation photons allow for further background suppression.

In this work, we investigate a complementary realization of this inelastic channel in the Short-Baseline Near Detector (SBND), a liquid argon time projection chamber (LArTPC) detector at Fermilab~\cite{MicroBooNE:2015bmn, SBND:2025lha}. We consider inelastic scattering of dark matter from argon nuclei, which excites nuclear states that subsequently de-excite via the emission of MeV-scale photons. The relevant argon excited states have excitation energies of several MeV and predominantly decay through the isotropic emission of one or more gamma rays. In a LArTPC, these photons produce short-range Compton-scattered electrons that appear as spatially localized, isolated energy depositions commonly referred to as ``blips”~\cite{ArgoNeuT:2018tvi, Castiglioni:2020tsu, MicroBooNE:2024prh}. Recent advances in LArTPC reconstruction have demonstrated sensitivity to individual MeV-scale energy depositions, enabled by excellent spatial resolution and low-noise detector performance. Building on these developments, we show that SBND can achieve sensitivity to dark matter–induced nuclear de-excitation signals using isolated MeV-scale electromagnetic activity. We consider both the ongoing target-mode operation of SBND and a proposed beam-dump configuration~\cite{SBND:2025lha}, establishing the ongoing SBND as a powerful probe of inelastic dark matter interactions at accelerator-based experiments.
Furthermore, this analysis, utilizing the state-of-the-art nuclear calculations, demonstrates the potential of MeV-scale nuclear de-excitation signatures in liquid-argon detectors for dark matter and supernova neutrino detections at DUNE.

{\bf{\emph{Dark Matter Model}.}} This analysis considers a fermionic dark matter particle $\chi$ that interacts with a U(1) gauge boson $A^{'}$ via a coupling $g_D$.
The gauge boson $A^{'}$ subsequently couples to Standard Model (SM) charged fermion $f$ with coupling $g_f$ in the vector-portal dark photon~\cite{Holdom:1985ag, Boehm:2003hm, Fayet:2004bw, Pospelov:2007mp, Arkani-Hamed:2008hhe, deNiverville:2016rqh} scenario, or in the leptophobic model~\cite{Batell:2014yra, Coloma:2015pih, Dror:2017ehi, Berlin:2018bsc, Boyarsky:2021moj, Gninenko:2014pea}.
The relevant Lagrangian is given by
\begin{equation}
\mathcal{L_{\text{int}}} \supset g_D \bar{\chi}\gamma^\mu \chi A^{'}_\mu + \sum_{f}g_f \bar{f}\gamma^\mu f A^{'}_\mu
\end{equation}

where, $g_f =
\begin{cases}
\varepsilon\, e\, Q_f, & \text{Vector-portal dark photon model}, \\
g_q\,\delta_{f,q}, & \text{Leptophobic model}.
\end{cases}$

Here, $\varepsilon$ and $g_q$ denote the couplings for the dark photon and leptophobic models, respectively. The parameter $\delta_{f,q}$ equals 1 when the fermions are quarks and 0 otherwise.

The simple dark photon–dark matter anomaly-free model, which utilizes photon-like couplings of the new gauge boson (associated with the new $U(1)$ symmetry), was originally proposed in the US Cosmic Vision: Community Report~\cite{Battaglieri:2017aum} as a conservative benchmark for presenting constraints in the dark matter parameter space. The main motivation for expressing the parameter space in terms of this minimal Lagrangian is that it enables a straightforward reinterpretation of experimental constraints within more complete or UV-complete models.

The U(1) gauge boson can acquire mass via Stueckelberg mechanism (see~\cite{Ruegg:2003ps} and references therein). However, if the $A'$ mass arises from spontaneous symmetry breaking, an associated Higgs-like scalar will also be present (e.g.,~\cite{Dutta:2019fxn}). This scalar can be produced through the proton bremsstrahlung process, via photons and electrons, as well as through three-body decays of pions (e.g., ~\cite{Dutta:2025sba}). Depending on how this scalar couples to dark matter, these additional production channels can enhance the DM flux.

The leptophobic model considered here to emphasize that the process described above is purely a quark-coupling--driven process, from production to detection, with the signal rate depending on $g_q^4 \alpha_D$. Only COHERENT and CCM, which utilize nuclear recoils, provide constraints directly in this parameter space in Refs.~\cite{CCM:2021yzc, COHERENT:2022pli}. In contrast, the constraints from LSND and MiniBooNE arise from processes proportional to $g_q^2 \epsilon_\ell^2 \alpha_D$, since they rely on electron-recoil signals, while NA64$(e,\mu)$ probes depend only on $\epsilon_{e,\mu}^2$ and therefore, do not provide competitive constraints in this parameter space. There exist anomaly-free models in the literature without direct electron/muon couplings but with direct quark couplings~\cite{Burk:2026mvk}, such as $U(1)_B$ models involving $N$ and new vector-like fermions under the SM gauge group~\cite{Carena:2004xs}, the $U(1)_{B-3L_i}$ class of models involving $N$~\cite{Bauer:2020itv}, and $U(1)_{T{3R}}$ models that couple only to right-chiral first-generation quarks and third-generation leptons~\cite{Dutta:2019fxn}.

\begin{figure*}[!htbp] 
    \centering
    \subfloat[]{
        \resizebox{0.2\textwidth}{!}{
            \begin{tikzpicture}
            \tikzset{every node/.style={font=\large}}
                \begin{feynman}
                    \vertex (a) at (-2,0) {$\pi^0/\eta$ };
                    \vertex (b) at (0,0);
                    \vertex (d) at (1.5,1.5) {$\gamma$};
                    \vertex (e) at (1.5,-1.5) {$\textcolor{violet}{A^{'}}$};
                    \diagram*{
                    (a) -- [scalar, very thick](b), (b) -- [fermion, very thick](d),
                    (b) -- [violet, photon, very thick](e),
                    };
                \end{feynman}
            \end{tikzpicture}
        }
    \label{fig:neutralmesondecay}
    }
    \hfill
    \subfloat[]{
        \resizebox{0.2\textwidth}{!}{
            \begin{tikzpicture}
            \tikzset{every node/.style={font=\large}}
                \begin{feynman}
                    \vertex (m) at (0,1);
                    \vertex (a) at (-2,1) {$p$};
                    \vertex (b) at (2, 1) {$p$};
                    \vertex (e) at (1, 1);
                    \vertex (f) at (1.5, 2.5) {$A^{'}$};
                    \vertex (c) at (-2, -1.3) {$N$};
                    \vertex (d) at (2, -1.3) {$N$};
                    \vertex (n) at (0, -1.3);
                    \node at (0.3, 0.0) {$\gamma$};
                    \diagram*{
                    (a) -- [fermion, very thick] (m) -- [fermion, very thick] (b),
                    (m) -- [boson, very thick](n), (e) -- [violet, boson, very thick](f),
                    (c) -- [double,double distance=0.5ex,thick,with arrow=0.5,arrow size=0.2em](n) -- [double,double distance=0.5ex,thick,with arrow=0.5,arrow size=0.2em](d)
                    };
                \end{feynman}
            \end{tikzpicture}
        }
    \label{fig:protonbrem}
    }
    \hfill
    \subfloat[]{
        \resizebox{0.2\textwidth}{!}{
            \begin{tikzpicture}
            \tikzset{every node/.style={font=\large}}
                \begin{feynman}
                    \vertex (a) at (0.0, 0.0);
                    \vertex (b) at (-2.0, 0.0) {$\textcolor{violet}{A^{'}}$};
                    \vertex (c) at (1.5, 1.5) {$\textcolor{blue}{\chi}$};
                    \vertex (d) at (1.5, -1.5) {$\textcolor{blue}{\bar{\chi}}$};
                    \diagram*{
                    (b) -- [photon, violet, ultra thick](a),
                    (a) -- [fermion, blue, very thick](c),
                    (d) -- [fermion, blue, very thick](a)
                    };
                \end{feynman}
            \end{tikzpicture}
        }
    \label{fig:A'decaytoDM}
    }
    \captionsetup{justification=Justified, singlelinecheck=off}
    \caption{Feynman diagrams depicting (a) the production of $A^{'}$ via neutral meson two-body decay, (b) the production of $A^{'}$ through proton bremsstrahlung, and (c) the two-body decay of $A^{'}$ into a $\bar{\chi}$ and $\chi$ pair.}
    \label{fig:FeynDiagram}
\end{figure*}
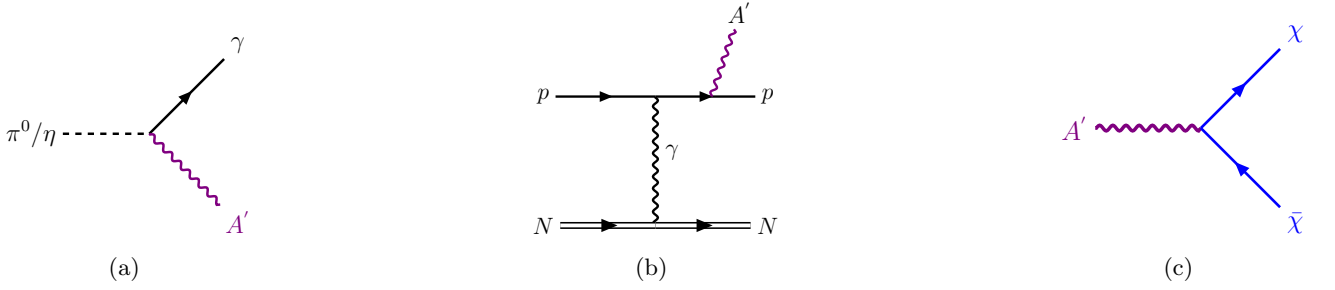

\begin{figure*}[!htbp]
    \centering
    \begin{subfigure}{0.48\textwidth} 
        \centering
        \includegraphics[scale=0.43]{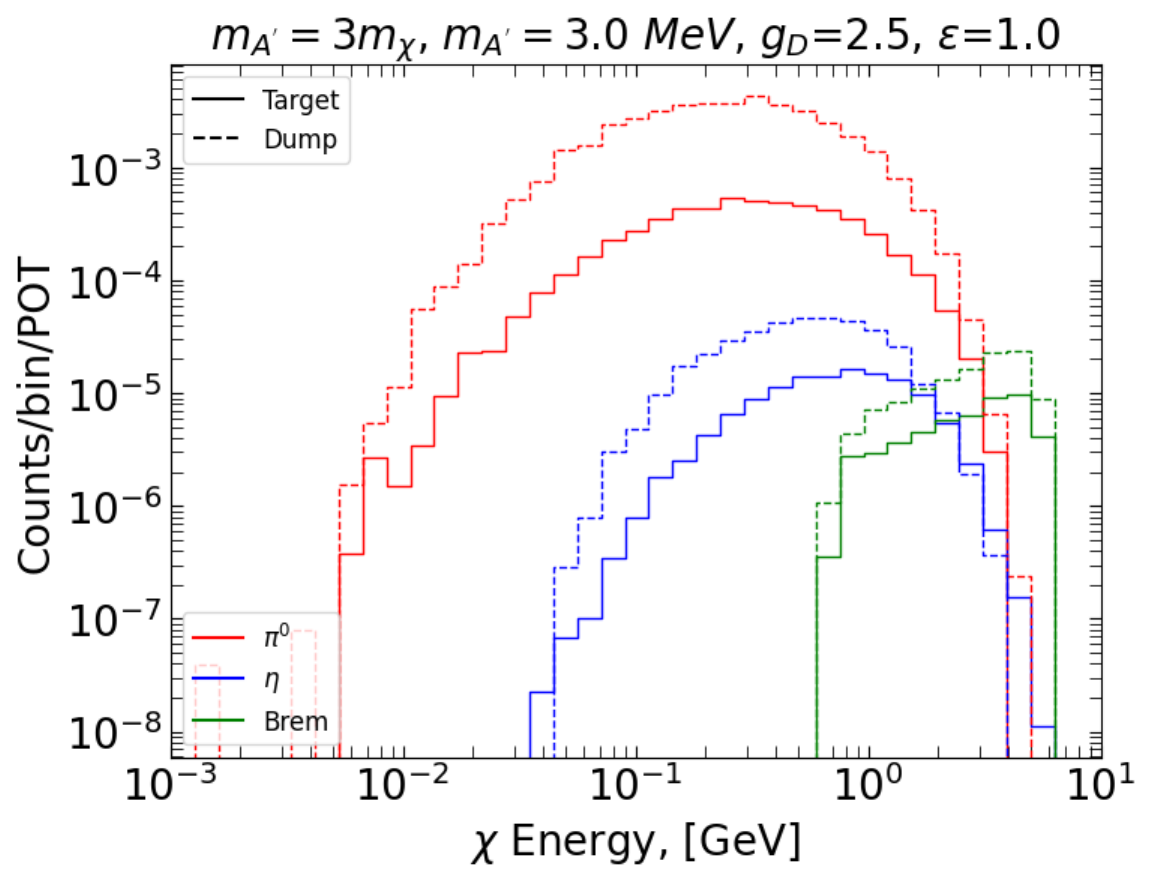}
        \caption{DM mass $m_\chi=1~$MeV}
        \label{fig:DMFlux1MeV}
    \end{subfigure}
    \begin{subfigure}{0.48\textwidth} 
        \centering
        \includegraphics[scale=0.43]{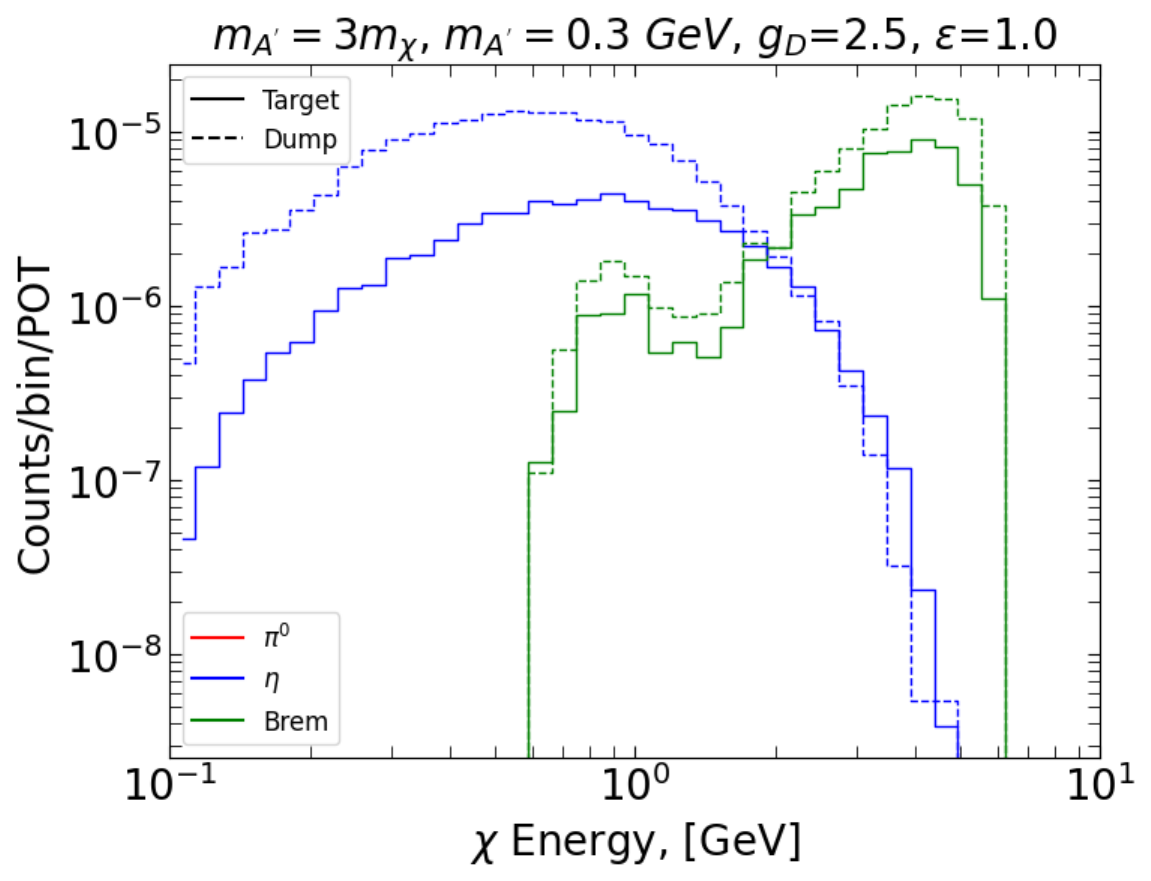}
        \caption{DM mass $m_\chi=100~$MeV}
        \label{fig:DMFlux100MeV}
    \end{subfigure}
    \captionsetup{justification=Justified, singlelinecheck=false}   
    \caption{DM fluxes with masses of (a) 1 MeV and (b) 100 MeV are shown at the front face of the SBND. Red, blue, and green lines correspond to DM production from $\pi^0$, $\eta$, and bremsstrahlung processes, respectively. Solid and dashed lines denote parent particles produced at the target and the dump, respectively. Uniform binning is applied to the x-axis in log-space, representing total DM energy. Thus, the y-axis shows counts per bin rather than counts per bin width. The total number of DM particles is calculated by summing the values for each bin, without multiplying by the bin widths. The red line ($\pi^0$ decay) is absent in (b) because $A^{'}$ has a greater mass than $\pi^0$, which prevents $\pi^0$ from decaying to produce $A^{'}$ on-shell. For these plots we consider, $\alpha_D=\frac{g_D^2}{4\pi}=0.5$, $m_{A^{'}}=3m_\chi$, and $\varepsilon=1.0$.}
    \label{fig:DMFlux}
\end{figure*}

{\bf{\emph{Dark Matter Production at SBND}.}} SBND is a $4 \times 4 \times 5$ $m^3$ LArTPC detector located 110 m downstream of the 70 cm-long $^9$Be Booster Neutrino Beam (BNB) target, which receives 8 GeV protons~\cite{MicroBooNE:2015bmn, SBND:2025lha}. Simulations using \texttt{GEANT4}~\cite{Allison:2006ve, Allison:2016lfl, GEANT4:2002zbu} indicate that approximately $18\%$ of the protons on target (POT) pass through the target without interacting and subsequently strike an iron dump with dimensions $4 \times 4 \times 4.21$ $m^3$, situated 50 m downstream along the beamline. At both the target and the iron dump, these protons generate a variety of secondary particles, including $\pi^0$ and $\eta$ mesons. The SM neutral meson ($\pi^0$ and $\eta$) fluxes are simulated~\cite{Capozzi:2021nmp, Capozzi:2024pmh} in both the beryllium target and iron dump using the \texttt{GEANT4} simulation package~\cite{Allison:2006ve, Allison:2016lfl, GEANT4:2002zbu}. The U(1) gauge boson $A^{'}$ is produced through the decay of these neutral mesons via
$\pi^0/\eta \to \gamma + A'$ as illustrated in Fig.~\ref{fig:neutralmesondecay}. Additionally, protons can produce $A^{'}$ directly through the proton bremsstrahlung process: $p + N \to p + A' + N$, as shown in Fig.~\ref{fig:protonbrem}. Dark matter (DM) is subsequently produced via the two-body decay of $A^{'}$, as depicted in Fig.~\ref{fig:A'decaytoDM}. For this analysis, $\alpha_D=\frac{g_D^2}{4\pi}=0.5$ and $m_{A^{'}}=3m_\chi$ is considered, which ensures that any on-shell $A^{'}$ decays promptly to a $\bar{\chi}$ and $\chi$ pair. Since $g_D \gg g_f$, we can safely assume a $100\%$ branching ratio for this channel. Under this assumption, all the production channels of $A^{'}$ and, henceforth, its subsequent decay to $\bar{\chi}{\chi}$ depend only on $\epsilon^2$, and therefore the flux of DM reaching the detector is independent of $\alpha_D$. Fig.~\ref{fig:DMFlux} presents the DM flux from the various sources and their respective production locations.

\begin{table*}[!htbp]
    \centering
    \renewcommand{\arraystretch}{1.2}
     \begin{tabular}{ | c | c | c | c | c |} 
        \hline
         Configuration & POT at Target & POT at Dump & Target Distance (m) & Dump Distance (m) \\
        \hline
        Target Mode (ongoing) & $8.2\times10^{20}$ & $1.8\times10^{20}$ & 110 & 60\\ 
        \hline
        Dump Mode (proposed) & ... & $6.0\times10^{20}$ & ... & 60\\
        \hline
    \end{tabular}
    \caption{Summary of the SBND operational modes considered in this analysis, including the corresponding protons-on-target (POT) exposures and detector distances.}
    \label{tab:ExperimentalDetails}
\end{table*}

This study presents results for two distinct scenarios:
\begin{itemize}
\item Target Mode: In this configuration, the proton beam is directed onto the Be target, corresponding to a total of $10^{21}$ protons on target accumulated over three years of operation. Even in this target-mode configuration, approximately 18\% of the protons escape the target and subsequently strike the dump.
\item Dump Mode: In this proposed scenario, the proton beam strikes the iron dump directly, bypassing both the Be target and the magnetic horn. This configuration reduces the neutrino background by a factor of 50~\cite{Toups:2022knq}, resulting in a significantly cleaner signal. Furthermore, it approximately doubles the production of neutral mesons.
\end{itemize}

A summary of the two operational modes is provided in Table~\ref{tab:ExperimentalDetails}.

{\bf{\emph{Inelastic DM-nucleus Scattering Cross-section: Ab Initio}.}} The multipole expansion of inelastic DM-nucleus scattering cross section is~\cite{Dutta:2022tav}
\begin{widetext}
\begin{equation}
\label{eq:dmInelCS}
\begin{split}
    \frac{d\sigma^{DM}_{\rm inel}}{dE_r}& = \frac{2e^2\epsilon^2 g_D^2 {E^\prime}_\chi p_\chi^\prime}{p_\chi {p'}_\chi(2m_N E_r + m_{A'}^2 - \Delta E^2 )^2} \frac{m_N}{2\pi} \frac{4\pi}{2J+1} \Bigg\{
    \sum_{J\geqslant 1, spin} \left[ \frac{1}{2}(\vec{l} \cdot \vec{l}^* - l_3 l_3^*) \left(
        \bigl| \langle J_f ||\hat{\mathcal T}_J^{mag}||J_i \rangle \bigr|^2 + \bigl| \langle J_f ||\hat{\mathcal T}_J^{el}|| J_i \rangle \bigr|^2
    \right) \right] \\
    & +\sum_{J\geqslant 0, spin} \left[ l_0 l_0^* \, \bigl| \langle J_f ||\hat{\mathcal M}_J|| J_i\rangle \bigr|^2 + l_3 l_3^* \, \bigl| \langle J_f||\hat{\mathcal L}_J|| J_i\rangle \bigr|^2 - 2\, l_3 l_0^* Re\left( \, \langle J_f||\hat{\mathcal L}_J|| J_i\rangle \langle J_f ||\hat{\mathcal M}_J|| J_i \rangle ^*
    \right) \right]\Bigg \}
\end{split}
\end{equation}
\end{widetext}
where $E_r$ is nuclear recoil energy, $m_N$ and $J$ denote the nuclear mass and spin, $J_{i(f)}$ is the initial (final) nuclear state, $E_\chi'$ and $p_\chi'$ are the outgoing DM energy and momentum, and $\Delta E$ is the nuclear excitation energy. Both elastic and inelastic cross-sections expressions contain the same dependence on  $\alpha_D$, $e$ and $\epsilon$.

The DM current in the scenarios considered here is purely vector: $l_\mu=\bar \chi \gamma^\mu \chi$. The dark matter current terms $\sum\limits_{s_i,s_f} l_\mu l_\nu^*$ of Eq.~(\ref{eq:dmInelCS}) are given in Ref.~\cite{Dutta2023}.

The electroweak multipole operators  $\hat{\mathcal{M}}_{JM}, \hat{\mathcal{L}}_{JM}, \hat{\mathcal{T}}_{JM}^{\rm el}, \hat{\mathcal{T}}_{JM}^{\rm mag}$ are projections of the weak hadronic current, $\hat{\mathcal{J}}$, such that they can act on nuclear states with good angular momentum and parity. They are defined by
\begin{eqnarray*}
    \hat{\mathcal{M}}_{JM} &=& \hat{M}_{JM} + \hat{M}^5_{JM} \nonumber\\
        &=& \int d^3x [j_J(qx)Y_{JM}(\Omega_x)] \hat{\mathcal{J}}_0(x) \nonumber\\
    \hat{\mathcal{L}}_{JM} &=& \hat{L}_{JM} + \hat{L}_{JM}^5 \nonumber\\
        &=& \frac{i}{q} \int d^3x \left[ \nabla[j_J(qx)Y_{JM}(\Omega_x)] \right]\cdot \hat{\mathcal{J}}(x) \nonumber\\
    \hat{\mathcal{T}}_{JM}^{\rm el} &=& \hat{T}^{\rm el}_{JM}+\hat{T}_{JM}^{\rm el5}\nonumber\\
        &=& \frac{1}{q} \int d^3x [\nabla \times j_J(qx) \textbf{Y}^M_{JJ1}(\Omega_x)]\cdot \hat{\mathcal{J}}(x)\nonumber\\
\end{eqnarray*}
\begin{eqnarray}
\hat{\mathcal{T}}_{JM}^{\rm mag} &=& \hat{T}_{JM}^{\rm mag} + \hat{T}_{JM}^{\rm mag5} \nonumber\\
        &=& \int d^3x [j_J(qx) \textbf{Y}^M_{JJ1}(\Omega_x)]\cdot \hat{\mathcal{J}}(x)
    \label{eq:mops}
\end{eqnarray}
where $j_J(qx)$ are Bessel functions, $Y_{JM}(\Omega_x)$ are spherical harmonics and $\textbf{Y}^M_{JJ1}(\Omega_x)$ are vector spherical harmonics. The weak hadronic current has V-A structure $\hat{\mathcal{J}}_\mu=\hat{J}_\mu + \hat{J}_\mu^5$, allowing the operators to be split into components of natural and unnatural parity. The normal parity operators: $M_{JM}, L_{JM}, T^{el}_{JM},$ and $T_{JM}^{mag5}$ can only contribute to transitions with $\Delta \pi = (-1)^J$. Similarly only $M^5_{JM}, L^5_{JM}, T^{el5}_{JM},$ and $T_{JM}^{mag}$ can contribute to unnatural parity transitions $\Delta \pi = (-1)^{J+1}$. The V-A structure is needed for calculating the SM neutrino background, while the V part is used for calculating the DM scenarios.

\begin{figure*}[!htbp]
    \centering
    \includegraphics[width=\columnwidth]{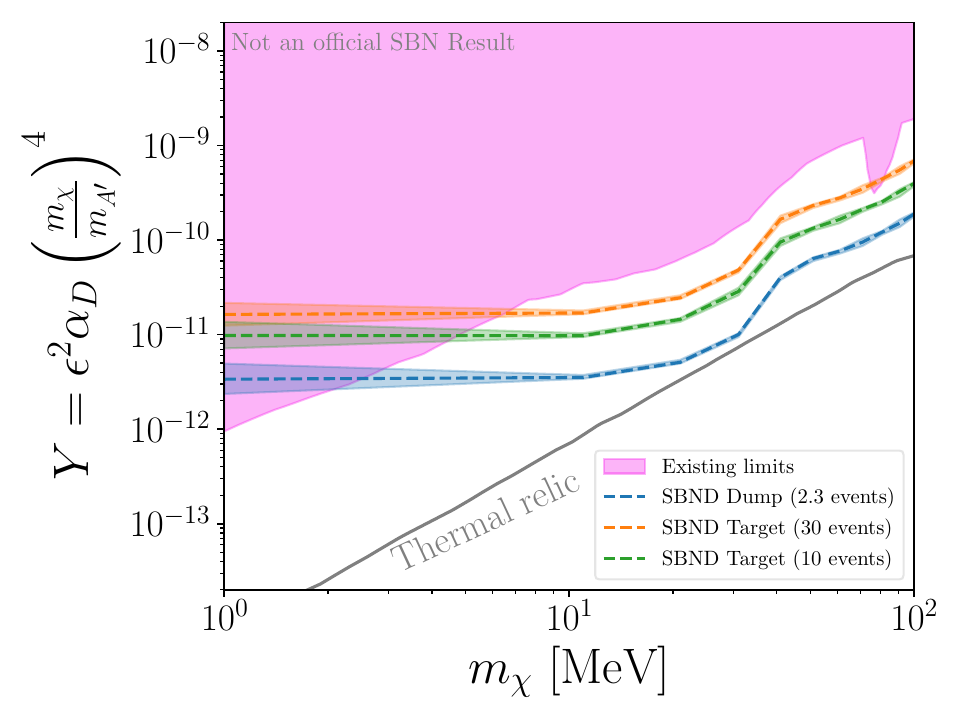} 
    \includegraphics[width=\columnwidth]{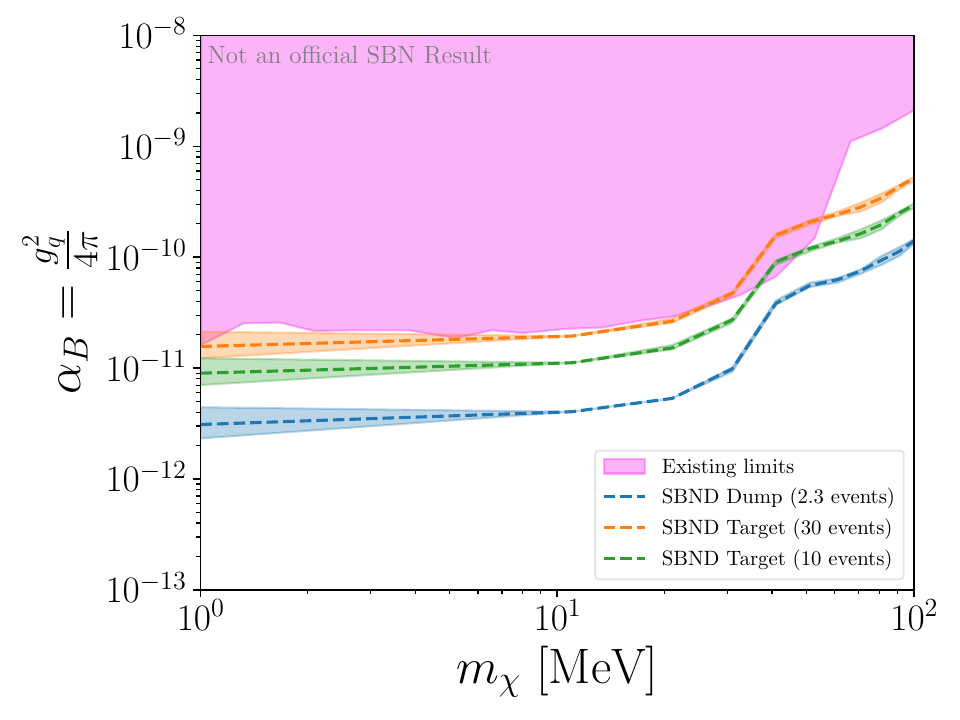}
    \captionsetup{justification=Justified, singlelinecheck=false}
    \caption{Dark matter exclusion bounds at SBND obtained using ab initio nuclear calculations for the vector-portal dark photon model (left) and the leptophobic dark matter model (right), where $\alpha_B = {g_q^2 \over 4\pi}$. The assumed numbers of signal events are indicated in the legend. The bands associated with the SBND lines represent the nuclear uncertainty quantified by different chiral interactions. For \enquote{SBND Target} lines, we use $8.2 \times 10^{20}$ POT on the Be target and the leaked $1.8 \times 10^{20}$ POT on the iron dump, as shown in Table~\ref {tab:ExperimentalDetails}. For \enquote{SBND Dump} mode lines, we use only $6.0 \times 10^{20}$ POT on the iron dump.}
    \label{fig:sen}
\end{figure*}

The phenomenological large-scale shell model (LSSM) has so far provided the most prominent calculations of the inelastic DM cross section. In the LSSM, however, operators must typically be adjusted (or ``quenched") to account for neglected many-body physics outside the chosen valence space and to improve agreement with data. While certain operators can be calibrated using Standard Model data, such as magnetic dipole (M1) transitions, these adjustments are insufficient to determine the full set of operators involved in DM scattering.

By contrast, ab initio nuclear calculations start from realistic nuclear forces and fundamental electroweak currents and solve the nuclear many-body problem without introducing phenomenological parameters. This first-principles approach yields nuclear form factors that are more physically robust and predictive. Over the past decade, ab initio nuclear theory has made rapid progress, significantly extending its mass reach and physics scope~\cite{hergert2020, Ekstrom2023fphy, hu2022a, ye2025}. Recently, ab initio methods have been applied to compute nuclear form factors for elastic dark matter–nucleus scattering in medium- and heavy-mass nuclei~\cite{hu2022b}. Extending these computations to inelastic scattering presents challenges: while elastic scattering requires only the ground-state wave function, inelastic calculations require computing wave functions for multiple excited states and the high-rank transition matrix elements between them. This substantially increases the computational cost. In this work, we extend a state-of-the-art ab initio method, valence-space in-medium similarity renormalization group (VS-IMSRG) \cite{tsukiyama2011,morris2015,hergert2016,stroberg2017,miyagi2020}, to accurately compute the inelastic DM-nucleus scattering cross section.

Our nuclear ab initio calculations begin with Hamiltonians expressed in the harmonic oscillator (HO) basis with frequency $\hbar \omega$, which we transform, along with the nuclear form factor operators, to the natural-orbital (NAT) basis, constructed from the perturbatively improved one-body density matrix  \cite{tichai2019,novario2020,hoppe2021,hu2022b}. 
We set $e=2n+l \leq (e_{\rm max}^{\rm NAT}=14)$, where $n$ and $l$ denote the radial and orbital angular momentum quantum numbers, respectively and impose an additional truncation $e_1+e_2+e_3 \leq (E_{\rm 3max}=22)$ for 3N  matrix element storage~\cite{Miyagi2022}. 
Next, we employ the Magnus formulation of the VS-IMSRG \cite{tsukiyama2011,morris2015,hergert2016,stroberg2017,miyagi2020} within a smaller subspace $e_{\rm max}\leq e_{\rm max}^{\rm NAT}$ to decouple a valence-space Hamiltonian, using ensemble normal ordering to approximately capture 3N forces between valence nucleons~\cite{stroberg2017}. 
The electroweak multipole operators are consistently transformed to generate effective valence-space operators, truncated at the two-body level, the IMSRG(2) approximation. The model space within $e_{\rm max}^{\rm NAT}=14$, $e_{\rm max}=8$ and $\hbar\omega=16$ MeV was demonstrated in Ref.~\cite{hu2022b} to be sufficient for obtaining converged nuclear form factors form light nuclei to heavy nuclei such as $^{129}$Xe.
We employ \{1s$_{1/2}$, 0d$_{3/2}$, 0f$_{7/2}$, 1p$_{3/2}$\} valence space with $^{28}$Si core. We then perform exact diagonalizations with the KSHELL code~\cite{shimizu2019}. In this work, three commonly used chiral two- and three-nucleon interactions are employed: 1.8/2.0(EM)~\cite{hebeler2011,simonis2017}, $\Delta$NNLO$_{\rm GO}$(394)~\cite{jiang2020}, N$^3$LO$_{\rm Texas}$ \cite{Hu2025a}.

{\bf{\emph{Sensitivity}.}} For argon nuclei, we consider de-excitation photons with energies up to 18 MeV, which includes contributions from nuclear excited states ranging from $1^+$ to $4^+$ and $0^-$ to $4^-$. Among all these states, the dominant contributions emerge from the $1^+$ states located roughly between 4.5 and 10 MeV, as well as the first $2^+$ state at an excitation energy of approximately 1.5 MeV. The excitation energy of the $2^+_1$ state calculated by the ab initio VS-IMSRG using three chiral interactions ranges from 1.32 to 1.98 MeV, in good agreement with the experimental value of 1.46 MeV. The ab initio $1^+$ excitation energies also reproduce the experimental data within about 0.5 MeV.

The LSSM calculation is valid for the $1^+$ states when the operators are rescaled to reproduce experimental M1 transition strengths, while the ab initio calculations, without such adjustments, successfully reproduce the available experimental data~\cite{Tornow:2022kmo}.
Inelastic DM-nucleus scattering not only produces MeV-scale de-excitation photons, but also produces a small nuclear recoil of keV-scale. 
The cascade MeV-scale energy will diminish the keV-scale nuclear recoil energy, and therefore, the contribution of the latter can be ignored.

The ability to reconstruct such MeV-scale electromagnetic activity in LArTPC detectors has been demonstrated by the ArgoNeuT experiment through the observation of nuclear de-excitation \(\gamma\) rays~\cite{ArgoNeuT:2018tvi}, and more recently by the MicroBooNE experiment using ambient radiogenic and cosmogenic activity to validate MeV-scale blip reconstruction, energy calibration, and background characterization~\cite{MicroBooNE:2024prh}. We investigate the DM parameter space fixing the mass ratio $m_{A'}/m_\chi = 3$ and $\alpha_D=\frac{g_D^2}{4\pi}=0.5$.
The resulting exclusion bounds for the experiments are shown in Fig.~\ref{fig:sen}. We show 2.3 events line for the SBND dump mode proposal, while 10 and 30 events lines for the target mode run.
These target events are obtained from 90\% CL assuming no SM background (dump mode) and estimated neutrino background (target mode).
The shaded region shows the existing limits from various elastic scattering searches~\cite{COHERENT:2021pvd,LSND:2001PRD,CCM:2021leg,MiniBoone:2018PRD,NA64:Gninenko_2021,E137:1988PRD,PhysRevD.104.L091701}.

For these representative dark photon-dark matter models, the relic abundance line, along with the MiniBooNE and LSND constraints and the SBND signal line, shifts identically upward or downward in terms of the $Y$ values, which $\propto\epsilon^2\alpha_D$. In contrast, the NA64 constraint, which is based on missing-energy searches rather than dark matter appearance type measurements, is independent of $\alpha_D$, as discussed in Ref.~\cite{NA64:2025ddk}. The choice of this $\alpha_D$ value was proposed in the US cosmic vision: community study~\cite{Battaglieri:2017aum} to obtain a conservative constraint in the parameter space. Although the couplings between the dark photon and dark matter are assumed to be large, the couplings to the visible sector particles remain much smaller. The large couplings, or the associated higher-order corrections, do not significantly modify any of the constraint lines shown in the plots, since the value of $\alpha_D$ can be rescaled to accommodate such higher-order corrections.

The thermal relic line corresponds to the standard freeze-out scenario, where dark matter was once in thermal equilibrium with the Standard Model plasma and reproduces the observed dark matter abundance~\cite{Planck:2018vyg}. If dark matter is thermal in origin, only parameters near this line are favored. However, non-thermal production mechanisms, such as freeze-in, particle decays, or asymmetric dark matter, can also generate the observed abundance~\cite{Cirelli:2024ssz}, allowing viable parameter space both above and below the thermal relic line. Therefore, a signal in these experiments could provide important information about the cosmological origin of dark matter.

The SN1987A constraint for this dark photon–dark matter model with $\alpha_D = 0.5$ has been calculated in Ref.~\cite{Chang:2018rso}. The resulting constraint corresponding to values of $Y \lesssim 10^{-16}$ represents the excluded part of the parameter space. 
The BBN constraint for this class of models can also be relevant for dark matter masses below $\sim 10,\mathrm{MeV}$, as shown in Ref.~\cite{Giovanetti:2021izc}. 

The relic abundance for $U(1)_B$-type leptophobic models has been computed in the literature, e.g., Ref.~\cite{COHERENT:2022pli}, showing that thermal dark matter in this scenario is largely ruled out. In contrast, the $U(1)_{T{3R}}$ model involves an additional scalar~\cite{Dutta:2019fxn}, which makes the relic abundance constraint less restrictive. 

{\bf{\emph{Backgrounds}.}} The signal considered here, MeV-scale nuclear de-excitation photons from inelastic dark matter scattering on argon, overlaps with several sources of low-energy electromagnetic activity in SBND.
The dominant irreducible physics background arises from neutral-current neutrino-argon inelastic scattering that excites the nucleus and produces de-excitation photons in the same MeV energy range as $\chi$ signal.
Using the expected BNB exposure in target mode, we estimate a total of \(\sim 235\) neutrino-induced inelastic events in SBND~\cite{MicroBooNE:2015bmn}.
In addition, neutrino interactions produce final-state neutrons that can scatter inelastically on surrounding argon nuclei, generating secondary de-excitation photons.
Compared to prompt nuclear de-excitation at the primary interaction vertex, such neutron-induced photons are typically spatially displaced and delayed, providing handles based on topology, multiplicity, and spatial distributions~\cite{ArgoNeuT:2018tvi}.
Non-electromagnetic de-excitation (eg, neutron, proton, $\alpha$ emissions) is not considered in this study as they have distinct spectra from photons.

Beyond neutrino-induced activity, beam-related neutrons constitute an important background class for MeV-scale searches near the BNB. These include neutrons produced at the target or dump that enter the detector volume, as well as two delayed, beam-correlated components identified in the ANNIE hall: ``sky-shine'' neutrons produced in the dump that scatter in air and enter the detector predominantly from above, and ``dirt'' neutrons produced by neutrino interactions in upstream material that enter from the beam direction~\cite{ANNIE:2019azm}. Measurements indicate that these neutron backgrounds are delayed relative to the prompt neutrino component and strongly dependent on detector position, with the highest rates near the top of the detector volume and a rapid reduction with depth~\cite{ANNIE:2019azm}. In SBND, such backgrounds can, in principle, be mitigated through fiducialization, spatial selections that suppress activity near the upstream and top regions, and timing selections relative to the beam spill. The narrow BNB spill structure further enables strong suppression of steady-state backgrounds, with residual beam-uncorrelated contributions constrained using beam-off data.

Additional non-beam backgrounds include cosmogenic activity, ambient \(\gamma\)-ray radiation from detector materials and the surrounding environment, internal radioactivity (notably \({}^{39}\)Ar), and electronics noise. These sources populate the MeV-scale regime and can produce isolated low-energy depositions that resemble photon-induced ``blips.'' Recent measurements by the MicroBooNE collaboration demonstrate that large liquid-argon time projection chambers can robustly reconstruct isolated MeV-scale electromagnetic activity with controlled energy calibration, resolution, and efficiency using ambient radiogenic and cosmogenic sources~\cite{MicroBooNE:2024prh}. These studies establish that data-driven background templates, beam-off samples, and topology-based selections can effectively characterize and constrain such low-energy backgrounds, providing a validated experimental framework for MeV-scale blip analyses in SBND-like detectors.

A dedicated timing- and topology-based background analysis incorporating the expected dark matter time-of-flight relative to prompt neutrino interactions, as well as spatial and multiplicity information, lies beyond the scope of this work; we therefore present sensitivities in terms of the expected number of signal events. For target-mode running, the 30-event contour corresponds approximately to \(3\sigma\) sensitivity when the estimated neutrino background is included, while we expect that a dedicated background analysis could plausibly extend sensitivity toward the 10-event level. For the proposed beam-dump configuration, the neutrino flux is suppressed by more than an order of magnitude relative to the target mode, substantially reducing neutrino-induced nuclear excitation backgrounds. We therefore present dump-mode sensitivities under the simplifying assumption of a background-free search, noting that residual contributions from beam-related neutrons and detector backgrounds can be constrained using timing, fiducialization, and beam-off data.

{\bf{\emph{Conclusion}.}} In conclusion, we have investigated blip events at SBND as a novel probe of light dark matter produced via the decays of light vector mediators and subsequently scattering inelastically from argon nuclei within the detector. We considered both dark photon–type and leptophobic models. For each scenario, we employed state-of-the-art ab initio nuclear calculations to accurately determine dark matter–nucleus inelastic scattering rates. For the first time, we systematically included all relevant nuclear excited states with energies up to 18 MeV and found that the $1^+$ states and the first $2^+$ state provide the dominant contributions to the signal. 
This approach enables robust signal predictions and improved background discrimination. Accounting for both neutrino- and neutron-induced background processes, we demonstrate that previously unexplored regions of parameter space in both models can be probed at SBND. These results highlight the potential of MeV-scale nuclear de-excitation signatures in liquid-argon detectors as a powerful avenue for dark matter and supernova neutrino detection at DUNE.

{\bf{\emph{Acknowledgment}.}} We would like to thank Steven Gardiner for helpful comments on the manuscript,  Danyan Pan, Wen Luo, and Shutong Zhang for helpful discussions on TALYS calculations, and Jayden Newstead for helpful discussions on ab initio calculations. B.~Dutta, D.~Goswami, and W.-C. Huang are supported by the U.S. Department of Energy (DOE) Grant No. DE-SC0010813. B.~S. Hu is supported by the Cyclotron Institute at Texas A\&M University. This manuscript has been authored by FermiForward Discovery Group, LLC under Contract No. 89243024CSC000002 with the U.S. Department of Energy, Office of Science, Office of High Energy Physics.\\
Note: The work and conclusions presented in this publication are not to be considered as results from the SBN Collaboration.

\bibliography{refs}

\end{document}